

\magnification=1200
\def\i#1{\item{#1}}

\def\b{$$\eqalignno}

\def\to{\rightarrow}
\def\.{\!\cdot\!}
\def\l{\ell}
\def\h{{1\over 2}}

\def\p{\partial}

\def\r2{\sqrt{2}}
\def\bk#1{\langle#1\negthinspace\rangle}
\def\ket#1{|#1\rangle}
\def\bra#1{\langle#1\negthinspace\negthinspace|}

\def\Tr{{\rm Tr}}
\def\){\right)}
\def\({\left(}
\def\[{\left[}
\def\]{\right]}
\def\.{\cdot}
\font\ninerm=cmr8

\nopagenumbers
\hsize=6.0truein
\vsize=8.5truein
\parindent=1.5pc
\baselineskip=10truept

\input epsf.tex
\font\small=cmr10
\font\bold=cmbx10

\rightline{McGill 94-20}
\rightline{hep-ph/9406388}
\rightline{(June, 1994)}
\bigskip
\centerline{{\bf STRING-ORGANIZED FIELD THEORY}
\footnote{$^*$}
{\ninerm Based on a talk given at the MRST meeting at McGill University,
May 11--13, 1994.}
}

\vglue 7truept
\vglue 1.0truecm
\centerline{  C.S. LAM
\footnote{$^\dagger$}
{\ninerm  Email: Lam@physics.mcgill.ca}}
\baselineskip=13truept
\centerline{\it Department of Physics, McGill University}
\baselineskip=12truept
\centerline{\it 3600 University St., Montreal, Qu\'ebec, Canada H3A 2T8}
\vglue  0.8 cm
\centerline{  ABSTRACT}
\vglue 0.3 cm
  {\rightskip=3 pc
 \leftskip=3 pc
 \baselineskip=10 truept
 \noindent
\ninerm
A low energy string theory should reduce to an ordinary quantum
field theory, but in reality the structures of the two are
so different as to make the equivalence obscure.
The string formalism is more symmetrical between the spacetime
and the internal degrees of freedom, thus resulting in considerable
simplification in practical calculations and novel insights in
theoretical understandings.
 We review here how tree or multiloop field-theoretical
diagrams can be organized in a string-like manner
 to take advantage of
this computational and conceptual simplicity.
\vglue 0.8truecm }
\line{\bf 1. Introduction \hfil}
\vglue 0.2 cm
\baselineskip=14truept

At present energies much less than $10^{19}$ GeV,
excited levels of a superstring cannot be reached, so a
superstring scattering amplitude ought not be distinguishable
from one
obtained from a massless quantum field theory (QFT). Yet the
formalism of a string theory is so vastly different from
a QFT that this equivalence is not at
all obvious. Specifically, for a superstring theory,
 (1) the fundamental dynamical
variables consist of the spacetime
 $x^\mu(\sigma,\tau)$ and the
internal   $\psi^i(\sigma,\tau)$ fields, all as functions
of the worldsheet coordinates $\sigma$ and $\tau$; (2)
these variables propagate as independent
free fields throughout the worldsheet, in a
manner dependent on the topology but not on
the geometry of the worldsheet
(reparametrization and conformal invariance);
(3) an external photon of momentum
$p$ and wave function $\epsilon_\mu(p)$ is inserted into
the string through a vertex operator $\epsilon(p)\.
[\p_\tau x (\sigma,\tau)]\exp[ip\.x(\sigma,\tau)]$, a form which is
fixed by conformal invariance; (4) conformal invariance leads to local
(Veneziano) duality$^1$ of the scattering amplitude, which enables
 a scattering process to be  described by
very few string diagrams. In particular, for elastic
scattering in the tree approximation, one string
diagram (the Veneziano amplitude)
gives rise simultaneously to all
 the $s$-channel and the
$u$-channel exchanges.

In constrast, in QFT,
(1$'$) the fundamental dynamical
variables are  fields $\psi^i(x)$ of the four-dimensional
spacetime coordinates $x^\mu$; (2$'$) these fields propagate
freely only between vertices, where interactions take place; (3$'$)
external photons are inserted into  a Feynman diagram through the
photon operator $ \epsilon(p)\.A (x)\exp[ip\.x]$;
 (4$'$) there are
many distinct Feynman diagrams contributing to a scattering
amplitude. For elastic scattering, $s$- and $u$-channel exchanges
are given by different diagrams that must be added up together.

The purpose of this talk is to discuss how QFT can be reformulated
in a string-like manner.
There are two advantages to reformulate QFT this way.
On the practical side, it frees the spacetime, spin,
and  color variables to  propagate independently (properties
(1) and (2) of string) so that the scattering amplitude can be
decomposed into sums of factorized forms much easier to handle.
This for example makes it possible for the spinor helicity
technique$^2$ to be applied to loop graphs$^3$. The string-like external photon
vertex (3) allows gauge invariance to be looked at in a  new way$^4$
that simplifies calculations. If duality (4) can be implemented
in QFT as well, then one has effectively a way of summing a number
of diagrams. All of these make it possible to compute processes
that are difficult or impossible to do in the usual way$^{5-9}$.
On the theoretical
side, a string-like reformulation of QFT allows new insights
to be obtained by studying gauge invariance in
 gauge and gravitational theories from a new perspective.
 It might even help the calculation of
multiloop string amplitudes by knowing how to do it for
the simpler but similarly-formulated multiloop QFT amplitudes.

For that to happen, a proper-time variable
$\tau $ must be introduced
into QFT. It is not necessary to introduce the other string
variable $\sigma $ because string excitations are absent so this
variable is effectively frozen. Proper time is introduced by the
Schwinger representation for scalar propagators, for the
Schwinger proper time $\alpha _r$ for a propagator $r=(ij)$ joining
vertices $j$ to $i$ is equal to the proper-time difference
$|\tau _i-\tau _j|$.

\vglue 0.6cm
\line{\bf 2. Spacetime Flow \hfil}
\vglue 0.4cm
We shall assume all particles involved to be massless.
By making the Schwinger proper-representation for
the denominator of every propagator
$${1\over q^2+i\epsilon}=-i\int_0^\infty d\alpha\exp(i\alpha q^2)\ ,
\eqno(2.1)$$
loop-momentum integrations of every QFT amplitude can be carried
out. The amplitude for a $\l$-loop Feynman diagram in $d$-dimensional space
with $N$ internal lines of momenta $q_r$ and
$n$ vertices  with outgoing momenta $p_i$ is then$^{10}$
$$ \eqalignno{
A&=\int [D \alpha ]\Delta (\alpha )^{-d/2}S(q,p)\exp[iP]\
,&(2.2)\cr
\int [D  \alpha ]\equiv &\[{(- i)^{d/2}\mu^\epsilon
\over  (4 \pi)^{d/2}}\]^{\l }i^{N}
\int _{0}^{\infty }(\prod _{r=1}^{N}d\alpha _{r})\ ,&\cr
P&=\sum_{r=1}^N\alpha_rq_r^2\equiv\sum_{i,j=1}^nZ_{ij}(\alpha)p_i\.p_j\
,&(2.3)\cr
&S(q,p)\equiv \sum _{k\ge 0}S_{k}(q,p)\ ,
&(2.4)}$$
where $S_{k}(q,p)$ is  obtained from $S_{0}(q,p)$ by
 contracting $k$ pairs of $q$'s in all possible ways, then
 summing over all the contracted results. The rule for contraction is:
$$q^{\mu }_{r}q^{\nu }_{s}\to -{i\over 2}H_{rs}(\alpha)g^{\mu \nu }
\equiv q^{\mu }_{r}\sqcup q^{\nu }_{s}\ .\eqno(2.5)$$
These formulas are
valid for tree diagrams as well. In that case $\Delta=1$
and $H_{rs}=0$.

If the Feynman diagram is regarded as an electric circuit with
branch resistances $\alpha _r$ and
outgoing external currents $p_i$,
then the quantity $q_r$ in (2.4) and (2.5) becomes the current flowing
through the $r$th internal line,  $P$ in (2.3) equals to the power
consumed in the circuit, and $Z_{ij}$ is the impedance matrix
of the network. The functions $\Delta$ and $H_{rs}$ can
be attached circuit interpretations as well. In this way all the
quantities in the amplitude can be expressed as electric circuit
quantities. There are explicit graphical rules$^{3,10,11}$ to obtain each of
them directly from the Feynman diagram but we will not discuss them
here. With the Schwinger parameters $\alpha _r$ interpreted as
proper-time differences, the spacetime flow $x^\mu (\tau )$
anticipated in the string theory is now given by the flow
of currents through the circuit.
Color and spin are contained exclusively in the factor $S(q,p)$;
 their flows will be discussed in the next two sections.

\vglue 0.6cm
\line{\bf 3. Color Flow\hfil}
\vglue 0.4cm
Let $T^a\ (0\le a\le N^2-1)$ be the $U(N)$ color generators.
The color coupling factor for three octet objects is
$$f_{abc}=-i\Tr(T^aT^bT^c-T^aT^cT^b)\ ,\eqno(3.1)$$
and that for four octet objects is
$$f_{abe}f_{ecd}=(-i)^2\Tr(T^aT^bT^cT^d-T^bT^aT^cT^d+T^bT^aT^dT^c-
T^aT^bT^dT^c)\ .\eqno(3.2)$$
Instead of a single vertex with a color factor $f_{abc}$, one
can produce two color-oriented$^{3,11}$ (c-o) vertices as in Fig.~1A,B,
corresponding
to the two terms on the rhs of (3.1): the order in the trace
is followed clockwise in the diagrams.
Similarly, one has four
c-o vertices for the four-gluon coupling as shown in
Fig.~1C,D,E,F. These c-o vertices can be
obtaind from a single one by twisting or flipping some of the lines.

\vskip -10.0 cm
\centerline{\epsfxsize 4.7 truein \epsfbox {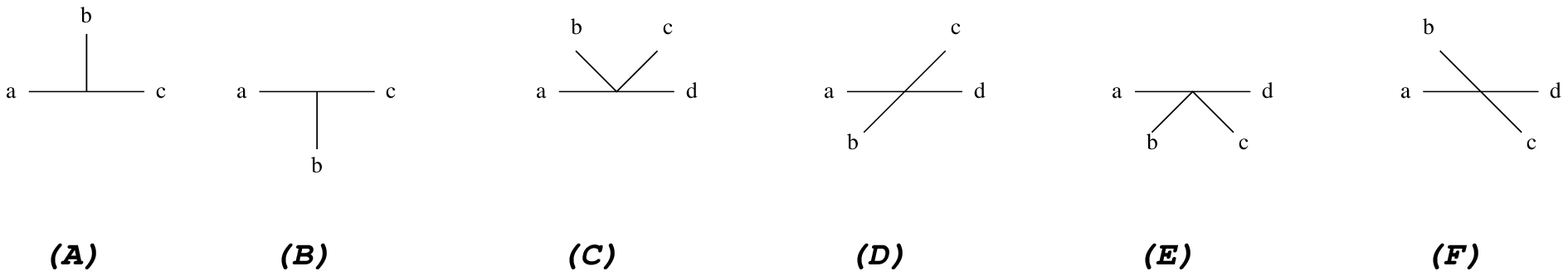}}
\nobreak
\vskip -1.0 cm\nobreak
{ \narrower\narrower\narrower\smallskip
{\bold Fig.~1:}\quad\quad{\small Color-oriented vertices. Read clockwise.}
\smallskip}
\vskip .2 cm

When strung together,
they produce c-o diagrams obtainable from a single Feynman
diagram by twisting the appropriate lines. So the c-o diagrams are
the QFT counterparts of the twisted diagrams in an open string theory.
Complete Feynman rules for c-o vertices can be
worked out from the usual Feynman rules$^3$.
Each c-o diagram obtained this way has a fixed
color factor, {\it i.e.,} a fixed color flow,
obtained by sewing
the $T^a$'s together using the $U(N)$ completeness relation
$$\sum_{a=0}^{N^2-1}(T^a)_{ij}(T^a)_{kl}=\delta _{il}\delta _{kj}\ .
\eqno(3.3)$$
The result$^{2,6}$
 is the generalization of the Chan-Paton factor$^{12}$
introduced for string theory. These color factors can be read off
directly from the
diagram$^{3,11}$ by following the external lines around, with the understanding
that
every fermion line is to be traversed once on top   and
every  gluon line is to
be traversed once on both sides. An open path results in a product of the
generators $T^a$, and a closed path results in the trace of such products.
For example, the color factor from Fig.~2 obtained this way is
$$\Tr\(T^{19}T^{20}\)\ T^{18}T^1T^2T^3\cdots T^{14}T^{15}T^{16}T^{17}\
.\eqno(3.4)$$
 \vskip -0  cm
\centerline{\epsfxsize 3.0 truein \epsfbox {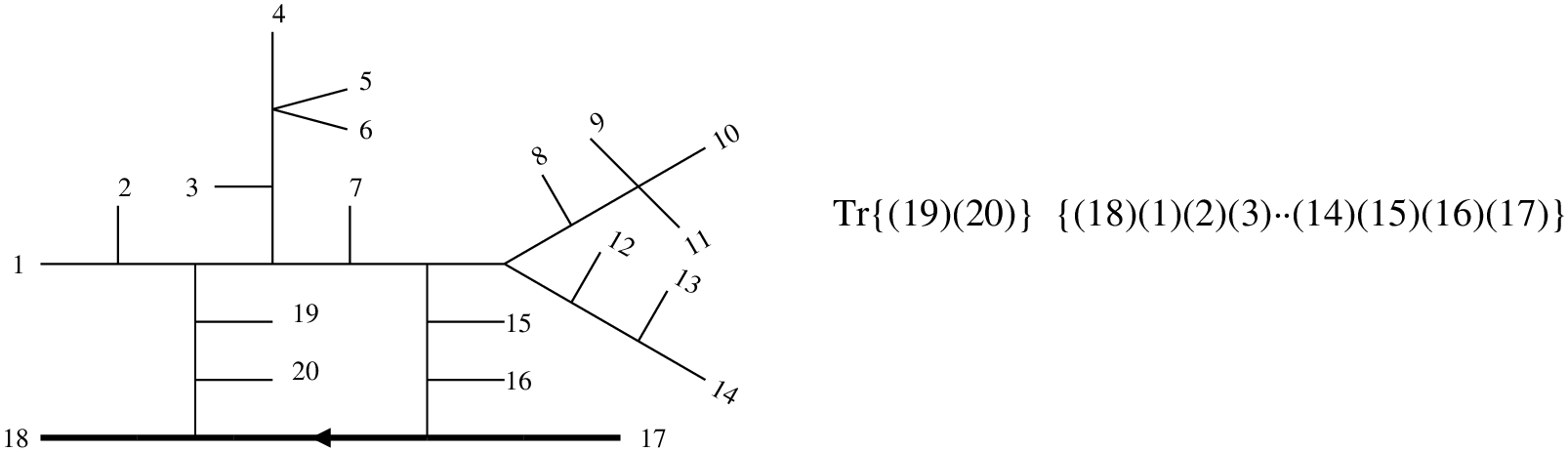}}
\nobreak
\vskip -5.0 cm\nobreak
{ \narrower\narrower\narrower\smallskip
{\bold Fig.~2:}\quad\quad{\small A color-oriented Feynman diagram.
The color factor is given by eq.~(3.4)}
\smallskip}
\vskip .2 cm

For $SU(N)$, the $a=0$ term of (3.3) has to be put on the right and
the result is more complicated.

\vglue 0.6cm
\line{\bf 4. Spin Flow \hfil}
\vglue 0.4cm
For massless fermions in a gauge theory, helicity and chirality are
conserved. This is the reason why fermion helicities can flow
freely across the vertex junctions to simulate the free flow (2)
found in a string theory. What about photons and gluons, whose
helicities are not conserved during interactions? These are spin-1
particles which kinematically can be thought of as composites of
two spin-$\h$ particles. Using this representation chirality
conservation can be used to bring about
a smooth spin flow across
the gauge particles as well.

Mathematically, if $\ket{p\pm}=u^{\pm}(p), \
\bra{p\pm}=\bar u^{\pm}(p)$ are the wave functions for
massless fermions,
then chirality conservation is reflected in the
equation $\bk{p_1\pm|p_2\pm}=0$, leaving only
$\bk{p_1+|p_2-}\equiv[p_1p_2]$ and
$\bk{p_1-|p_2+}\equiv\bk{p_1p_2}$ non-vanishing$^2$.
The kinematically-composite nature of the photon and the gluon
is mathematically represented by having their wave functions written as
$\epsilon_ \mu ^\pm(p,k)=\bk{p\pm|\gamma _ \mu |k\mp} /\r2
\bk{k\mp|p\pm}$, where $k$ is an arbitrary massless reference momentum
the different choice of which corresponds to a different gauge choice
of $\epsilon $. The $\gamma $-matrices
which mix the four spin channels can also be eliminated
because of chirality conservation. The relevant formulas are
$$\eqalignno{
\gamma p_i&=\ket{p_i+}\bra{p_i+}+\ket{p_i-}\bra{p_i-}&\cr
\bk{p_1+|\gamma^\mu|p_2+}\bk{p_3-|\gamma_\mu|p_4-}&=
2\bk{p_1+|p_4-}\bk{p_3-|p_2+}&\cr
\bk{p_1+|\gamma^\mu|p_2+}\bk{p_3+|\gamma_\mu|p_4+}&=
2\bk{p_1+|p_3-}\bk{p_4-|p_2+}\ .&(4.1)}$$
As a result, the spin part of the numerator factor $S(q,p)$ in
(2.2) and be written as sums of products of $\bk{p_ip_j}$ and $[p_ip_j]$.
A graphical rule can be worked out to read this off directly from
the Feynman diagram$^{3,11}$. For that purpose draw each photon and gluon
as a pair of quark lines as in Fig.~3, then connect together the
quark lines smoothly (according to a rule which we will not state here).
Then $S_0(q,p)$ (and similarly for $S_k(q,p)$)
can be obtained by following along the fermion lines, pausing
at external lines and internal {\it fermion} lines to
pair up the momenta. Note that along a given fermion line
the angular brackets $\bk{\cdots}$
and the square brackets $[\cdots]$ always appear alternately. Note also
that when offshell momenta appear, they are to be expressed as linear
combinations of $p_i$ (with $\alpha $-dependent coefficients) according
to the electric-circuit rules of Sec.~2. For example,
one can obtain from Fig.~3 in this way
$$S_0=e^4\.\bk{p_4q_2}[q_2p_3]\.[p_2q_1]\bk{q_1q_4}[q_4q_3]\bk{q_3p_1}\ ,
\eqno(4.2)$$
where
$$\bk{p_4q_2}[q_2p_3]\equiv\sum_{j=1}^4c_{2j}(\alpha )\bk{p_4p_j}[p_jp_3]
\eqno(4.3)$$
if $q_r=\sum_{i=1}^4 c_{ri}(\alpha )p_i$.
\vskip -9.9 cm
\centerline{\epsfxsize 5.0 truein \epsfbox {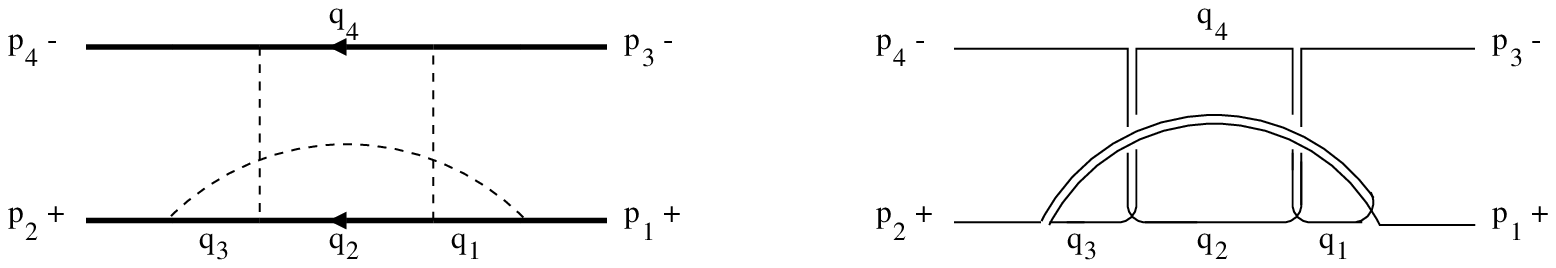}}
\nobreak
\vskip -2 cm\nobreak
{ \narrower\narrower\narrower\smallskip
{\bold Fig.~3:}\quad\quad{\small A Feynman diagram and its equivalent
diagram used to read off the spin factor in $S_0$. The result is given
in eqs.~(4.2) and (4.3).}
\smallskip}
\vglue 0.6cm
\line{\bf 5. String-like Vertex \hfil}
\vglue 0.4cm
The string-like vertex (3) can be obtained from the QFT vertex (3$'$)
by making use of a series of {\it differential circuit identities}$^{4,11}$,
some of which are shown below:
\b{
{\p P\over\p \alpha_s}&={\p \over\p \alpha_s}\(\sum_r\alpha_rq_r^2\)=q_s^2
\ ,&(5.1)\cr
{\p q_r\over\p \alpha_s}&=H_{rs}q_s\ ,&(5.2)\cr
{\p H_{rs}\over\p \alpha_t}&=H_{rt}H_{ts}\ .&(5.3)\cr}$$
Using these, one can change an external scalar QED vertex,
$V_a=e\epsilon(p_a)\.(q_{a'}+q_{a''})$,
where $p_a$ is the outgoing photon momentum and $q_{a''},q_{a'}$
are respectively the charged particle momenta pointing into and out
of the vertex,
into a string-like form
$V'_a=e \epsilon (p_a)\. \p_{ \tau_a} x(\tau_a )$.
As a result, the multiloop amplitude (2.2) for scalar QED
$$ A=\int [D \alpha ]\Delta (\alpha )^{-d/2}S(q,p)\exp[iP]\ ,$$
with $S=\sum_kS_k$, and
\b{
S_0&=\(\prod_{a=1}^{n_A} V_a\) S_0^{int}\ ,&(5.4)\cr
P&=\sum_{ij}p_i\.p_jZ_{ij}\ ,&(5.5)}$$
can be reduced to the string-like formula
$$ A=\int [D \alpha ]\Delta (\alpha )^{-d/2}S^{int}(q',p)\exp[iP']_{ML}\ ,
\eqno(5.6)$$
where $P'$ is obtained from $P$
by replacing every one of the $n_A$ $p_a$'s by
$p_a\to  p_a-ie \epsilon (p_a)\.\p_a$, where $\p_a=\p/\p \tau _a=
\p/\p \alpha _{a''}-\p/\p \tau _{a'}$. The subscript $ML$
in eq.~(5.6) tells us that the exponential should be expanded
and only terms multilinear in all $\epsilon _a$'s should be kept.
Similarly, $q_r'$ is obtained from $q_r$ by the same replacement,
except that $p_a\to  p_a-2ie \epsilon (p_a)\.\p_a$ should be used
if $a$ is one end of $r$.
\vglue 0.6cm
\line{\bf 6. Local Duality \hfil}
\vglue 0.4cm

String theory began with the discovery of the Veneziano model$^1$ for
two-particle scattering. In terms of the Mandelstam variables $s$ and
$u$, its amplitude is
$$A(s,u)=-B(-u,-s)=-\int_0^1x^{-u-1}
(1-x)^{-s-1}dx=-{\Gamma(-s)\Gamma(-u)\over \Gamma(-s-u)}\ .\eqno(6.1)$$
This is a meromorphic function with $s$-poles and $u$-poles located at
non-negative integers (in units of $[M_P\sim O(10^{19})$ GeV]$^2$),
but no simultaneous $s$- and $u$-channel poles. The amplitude can be
expanded either as a sum of $s$-channel poles,
represented purely by $s$-channel exchange Feynman diagrams, {\it or} a sum
of $u$-channel poles,   represented  purely by $u$-channel
exchange diagrams.
There is no need to {\it add} both the $s$-channel and the
$u$-channel diagrams, as is necessary in ordinary quantum field theories.
This is {\it duality} in string theory. What would be a suitable
definition of duality in QFT when there is not an infinite tower of
massive particles present?

At present energies when $|s|, |u|\ll 1$ (in units of $M_P^2$), only
massless poles in (6.1) contribute, and this yields
$$A(s,u)\simeq {1\over s} +{1\over u}\ .\eqno(6.2)$$
The $u$-channel pole comes from the divergence of the integral near $x=0$
when $u=0$, and the $s$-channel pole comes from the divergence of the
integral near $x=1$ when $s=0$.
In this form, the amplitude does not {\it appear} to be `dual' because
{\it both} the $s$-channel and the $u$-channel poles are {\it summed},
instead of having a {\it single} expression like (6.1), where only a sum
of the $s$-channel {\it or} a sum of the $u$-channel poles are present.
Nevertheless, appearances are deceiving, because (6.2) follows
mathematically from (6.1), which is dual.
In other words, the $u$-channel poles in (6.2)
can be formally obtained by an infinite sum of massive $s$-channel poles,
and (6.2) is as dual as it can be at low energies.

The amplitude for the   `Compton scattering' diagram of Fig.~4,
in a massless {\it scalar} QFT with interaction $\phi^*\phi A$,
is just given by (6.2) and therefore already dual. The only difference
one can point to is that $A$ in (6.1) is given by a {\it single}
integral, whereas (6.2) contains a sum of two terms.
We shall therefore define a {\it dual amplitude} in QFT as a sum
of Feynman diagrams which can be expressed as a single integral.
The question is whether QFT Feynman
diagrams can be summed up to give dual
amplitudes.
\vskip -2.8 cm
\centerline{\epsfxsize 4.5 truein \epsfbox {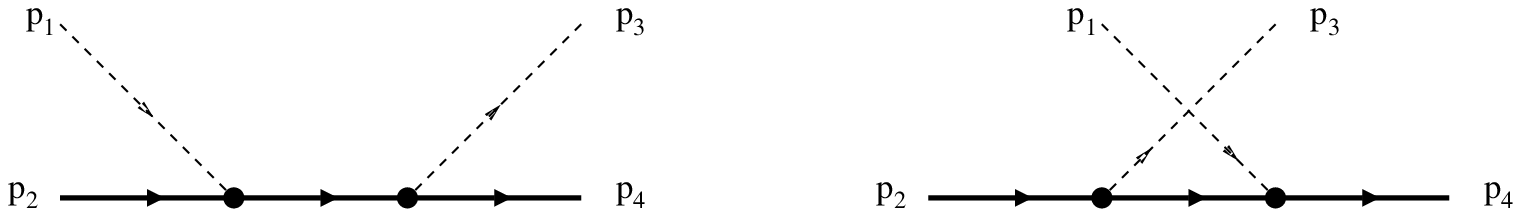}}
\nobreak
\vskip -8.2 cm\nobreak
{ \narrower\narrower\narrower\smallskip
{\bold Fig.~4:}\quad\quad{\small The $s$-channel and the $u$-channel
diagrams for Compton scattering.}
\smallskip}
\vskip .2 cm

The answer is yes for QED-like theories$^{13}$
but it is not yet clear for QCD. What makes QED special is that
the different diagrams in a gauge-invariant sum of diagrams  can
be distinguished by the orderings of their photon
vertices along the charged lines. If each vertex is assigned a
proper time $\tau $, then using (2.2) and expressing the $\alpha $'s
as differences of the $\tau $'s,  every diagram is
given by an integral over the $\tau $'s. The integration region
for different diagrams are different and non-overlapping.
 Symbolically, one can write the amplitude (2.2)
of a diagram to be
$$A=\int_R [D \tau ] T(\tau ,p)\exp[iP(\tau ,p)]\  ,\eqno(6.3)$$
where the integration region $R$ differs from diagram to diagram.
Since the different $R$'s do not overlapp, one can define an overall
function $T$ and an overall function $P$ in the union $C$ of the
$R$'s (which turns out to be a hypercube) to be equal to the
respective values of $T$ and $P$ in specific $R$ regions. In this
way, the dual sum of these gauge-invariant diagrams is simply
$$A_{sum}=\int_C [D \tau ] T(\tau ,p)\exp[iP(\tau ,p)]\  .\eqno(6.4)$$

How useful such a formal sum is depends on our ability to evaluate
the integral over the hypercube $C$. In general this is not easy
but in the eikonal approximation this can be done$^{13}$. Even when
(6.4) cannot be evaluated exactly it is still useful for two reasons.
It can be the starting point of an approximation (such as the
eikonal approximation) to sum diagrams. It can also be used
for mathematical manipulations ({\it e.g.}, the integration-by-parts
technique$^{6,8,9}$), effectively to shift gauge-dependent
 contributions between different
diagrams so that they do not appear even in the intermediate steps.

\vglue 0.6cm
\line{\bf 6. Summary \hfil}
\vglue 0.4cm

A string-theory scattering amplitude at an energy much below the
Planck mass should be the same as an appropriate massless QFT
amplitude, but on the surface they look different until one
uses the Schwinger-parameter representation. When this is done,
spacetime, spin, and color flows can be separated, spinor helicity
technique and string-like gauge vertices can be used to simplify
the calculations and to obtain novel insights into the structure of
gauge and gravitational theories. String-like local duality
can also be implemented in QED-like theories but its feasibility
is not yet clear for QCD.
\vglue 0.5cm
\line{\bf References \hfil}
\vglue 0.3cm
\def\i#1{\item{[#1.]}}
\def\npb#1{{\it Nucl.~Phys. }{\bf B#1}}
\def\plb#1{{\it Phys.~Lett. }{\bf #1B}}
\def\prl#1{{\it Phys.~Rev.~Lett. }{\bf B#1}}
\def\prd#1{{\it Phys.~Rev. }{\bf D#1}}

\i{1} G. Veneziano, {\it Nuovo Cimento} {\bf 57A} (1968), 190.
\i{2} P. De Causmaecker, R. Gastmans,
W. Troost, and T.T. Wu, \plb{105} (1981), 215;   \npb{291} (1987), 392;
 M.L. Mangano and S.J. Parke, {\it Phys. Rep.}  200 (1991), 301;
R. Gastmans and T.T. Wu, `The Ubiquitous Photon',
International Series of Monographs on
Physics, Vol.~80 (Clarendon Press, Oxford, 1990).
\i{3} C.S. Lam,  \npb{397} (1993), 143.
\i{4} C.S. Lam, \prd{48} (1993), 873.
\i{5} S. Parke and T. Taylor, {\it Phys. Rev. Lett.} 56 (1986),
2459.
\i{6}  Z. Bern and D.K. Kosower,
\prl{66} (1991), 1669; \npb{362} (1991), 389; \npb{379} (1992), 451.
\i{7} G. Mahlon, Fermilab preprint Fermilab-Pub-93/327-T (1993).
\i{8} Z. Bern, L. Dixon, and D.A. Kosower, \prl {70} (1993), 2677;
Z. Bern, L Dixon, D. C. Dunbar, and D. A.
Kosower,  SLAC-PUB-6415.
\i{9} M. Strassler, \npb{385} (1992) 145; SLAC-PUB 5978 (1992).
\i{10} C.S. Lam and J.P. Lebrun, {\it Nuovo Cimento} {\bf 59A} (1969), 397.
\i{11} C.S. Lam, {\it Can. J. Phys. } to appear.
\i{12} J. Paton and Chan H-M, \npb{10} (1969), 519.
\i{13} Y.J. Feng and C.S. Lam, to be published
\end